\documentclass[a4paper,12pt]{article}
\usepackage{amsmath,amssymb,amsfonts,epsfig,graphicx}
\usepackage{colordvi}
	\fboxrule 0.2mm 	\newcounter{comentario} 	

\title{Contribution of adiabatic phases to \\ 
                noncyclic evolution}

\author{
 M.T. Thomaz$^{1}\footnote{Corresponding author: mtt@if.uff.br}$,
 A.C. Aguiar Pinto$^{2}$ and M. Moutinho$^{2}$     \vspace{0.25cm} \\ 
\small\it $^{1}$Instituto de F\'{\i}sica, Universidade Federal Fluminense,\\ 
\small\it Av. Gal. Milton Tavares de Souza s/n$^{\textit o}$, 
CEP 24210-346, Niter\'oi-RJ, Brazil
\vspace{0.25cm}\\
\small\it $^{2}$Coordenadoria de F\'{\i}sica, Universidade Estadual de Mato Grosso do Sul,\\
\small\it Caixa Postal: 351, Cidade Universit\'aria de Dourados, CEP 79804-970, 
Dourados-MS,  Brazil.
}

\begin{document}

\maketitle

\begin{abstract}

We show that the difference of adiabatic phases,
that are basis-dependent, in  noncyclic evolution  
of non-degenerate quantum systems have to be taken 
into account to give the correct interference result in the 
calculation of physical quantities in states that are
a superposition of instantaneous eigenstates of energy. 
To verify the contribution of those adiabatic phases  
in the interference phenomena, we consider the spin-$1/2$ 
model coupled to a precessing external magnetic field.
In the model, the adiabatic phase increases in time up to reach the
difference of the Berry's phases of the model when the
external magnetic field completes a period. 
 
\end{abstract}

\vfill



\newpage

In 1928 Born and Fock\cite{fock} proofed the Adiabatic Theorem.
In a  quantum system with non-degenerate energy
spectrum, this theorem  says that if the system at $t=0$ 
is  in eigenstate of energy with quantum numbers $\{n\}$,
along an  adiabatic evolution it continues to be in 
an eigenstate of energy at time $t$  with the
same initial quantum numbers $\{n\}$. As  a consequence of this 
theorem, the vector state of the quantum system acquires an 
extra phase besides the dynamical phase. This extra phase 
is actually named geometric phase. Before the important
work by MV Berry in 1984\cite{berry} with cyclic adiabatic 
hamiltonian, this extra phase was realized to be dependent 
on the choice of the basis of instantaneous eigenstates 
of energy. This extra phase was considered non-physical
since it could be absorbed in the choice of the states in 
the instantaneous basis.\cite{messiah}.

In Ref. \cite{berry}, MV Berry showed that the adiabatic
phase acquired by the instantaneous eigenstates of energy,
after a closed evolution in the classical parameter space,
is physical due to its independence to the chosen basis to
describe the state vector at each instant. Since the 
publication of the Ref. \cite{berry}, the study
of Berry's phase has followed very interesting and broad
directions. More recently, the geometric phases have 
been proposed as a prototype for a quantum 
bit (qubit)\cite{leibfried,calarco,tian,imai}. In 1988 
Samuel and Bhandari\cite{samuel} generalized the 
geometric phase to noncyclic evolution. Many
others interesting papers appear to discuss those 
physical phases in noncyclic evolution in the classical 
parameter space\cite{gonzalo}. Experimental verification 
to the presence of those noncyclic geometric phases 
have been realized\cite{filipp}.

The interference effect is a keystone  in the linearity 
of the Quantum Mechanics. In the present letter we 
address to the question of the effect of  the adiabatic 
evolution on the phases in quantum systems leaves a 
physical trace in measurable quantities associated to
the noncyclic evolution of states described 
by a superposition of instantaneous eigenstates of 
energy. The same question was proposed in the nice
Ref. \cite{gonzalo}, but differently from them we do not
look for a physical noncyclic geometric phase.

\vspace{0.5cm}

Let us consider a time-dependent hamiltonian ${\bf H}(t)$
that evolves adiabatically. Following Ref. \cite{berry},
we leave open the possibility that this time dependence
comes from a set of classical parameters that we call
$\vec{R} (t)$ ($\vec{R} (t)\equiv (X_1 (t), X_2 (t), \cdots, X_m (t))$),
but we also include the possibility that the hamiltonian  
can have an explicit  time dependence. As a matter of 
simplification, we assume that the spectrum
of eigenvalues of ${\bf H}(t)$ is non-degenerate.

Let $\{ |\varphi_j; t \rangle, j= 1,2, \cdots\}$ be an instantaneous
basis of orthonormalized eigenstates of the energy

\begin{eqnarray} \label{1}  
{\bf H}(t) |\varphi_j; t \rangle = E_j (t) |\varphi_j; t \rangle, 
\end{eqnarray}

\noindent where 
$\langle \varphi_l; t | \varphi_j; t \rangle = \delta_{lj}$ 
and $l, j = 1, 2, \cdots$. 

We assume that the initial vector state is a superposition of 
$M$ eigenstates of energy at $t=0$,

\begin{eqnarray}  \label{2}
|\psi (0) \rangle = \sum_{j=1}^{M}  a_j \, |\varphi_j; 0 \rangle,
\end{eqnarray}

\noindent with $M>1$ and $\sum_{j=1}^{M} |a_j|^2 = 1$.

Applying the Adiabatic Theorem\cite{fock,messiah} to the
Schr\"odinger eq. of the adiabatic evolution of 
the initial vector state (\ref{2}), it gives,

\begin{subequations}

\begin{eqnarray}  \label{3a}
|\psi (t) \rangle = \sum_{j=1}^{M} a_j \, e^{i \gamma_j (t)}
   \, e^{-\frac{it}{\hbar} \langle E_j (t) \rangle }  \;
         |\varphi_j; t \rangle,
\end{eqnarray}   

\noindent where $\langle E_j (t) \rangle$ is the average energy
during the interval of time $t$,

\begin{eqnarray} \label{3c}
\langle E_j (t) \rangle \equiv \frac{1}{t} \, \int_0^t dt^{\prime} \, E_j (t^{\prime})
\end{eqnarray}

\noindent and $\gamma_j (t) \in \mathbb{R}$ is the adiabatic
phase,

\begin{eqnarray}   \label{3b}
   \gamma_j (t) = i \int_0^t  dt^{\prime} \, \langle \varphi_j; t^{\prime}| 
\left(\frac{d}{dt^{\prime}}| \varphi_j;t^{\prime} \rangle \right).
\end{eqnarray} 

\noindent It is well known that the adiabatic phase  (\ref{3b}) 
is non-physical. 

\end{subequations}

An obvious physical quantity to calculate from  the vector state 
(\ref{3a}) is the  density of probability to find the particle 
at position $\vec{x}$ at any instant $t$,

\begin{eqnarray}   \label{5}
|\psi (\vec{x}, t)|^2 &=& \sum_{j=1}^{M}  |a_j|^2 |\varphi_j (\vec{x}; t)|^2
        \nonumber  \\
%
%
&+&  \sum_{{j, l=1}  \atop {j \not= l}}^{M} a_j \, a_l^{*} 
  e^{ i [\gamma_j (t) - \gamma_l (t)]} \;\; 
    e^{- \frac{it}{\hbar} \left[\langle E_j(t) \rangle -  \langle E_l(t) \rangle\right]} \;\;
       \varphi_j (\vec{x};t) \varphi_l^{*} (\vec{x};t) ,
\end{eqnarray}

\noindent where $\psi (\vec{x}, t) = (\vec{x}| \psi(t)\rangle$ 
and $\varphi_j (\vec{x}; t) = (\vec{x}| \varphi_j; t\rangle$. 

The interference phenomenon comes from the terms on the 
second sum on the r.h.s. of eq.(\ref{5}). Each interference 
term depends only on the difference of adiabatic 
phases (\ref{3b}).

\vspace{0.5cm}

Let $\{|\Phi_j; t\rangle, \, j=1, 2, \cdots\}$ be another basis
of instantaneous eigenstates of energy,

\begin{eqnarray}  \label{6}
 | \Phi_j; t\rangle = e^{ i \alpha_j (t)} \; | \varphi_j; t \rangle ,
      \hspace{1cm} j= 1, 2, \cdots
\end{eqnarray}

\noindent and $\alpha_j (t) \in \mathbb{R}$ and its time-dependence
comes through $\vec{R} (t)$ and/or an explicit time dependence.
The initial state (\ref{2}) is rewritten in this basis as

\begin{subequations}

\begin{eqnarray} \label{7a}
|\psi (0)\rangle = \sum_{j=1}^{M} \tilde{a}_j | \Phi_j; 0\rangle.
\end{eqnarray}

Therefore

\begin{eqnarray}  \label{7b}
 \tilde{a}_j = a_j \, e^{-i \alpha_j(0)}  .
\end{eqnarray}

\end{subequations}

Written in the new basis, $|\psi (t)\rangle$ becomes

\begin{eqnarray}  \label{8}
|\psi(t)\rangle = \sum_{j=1}^{M} \tilde{a}_j \, e^{i \tilde{\gamma}_j (t)}  \;
    e^{-\frac{it}{\hbar} \langle E_j (t) \rangle} \; |\Phi_j; t\rangle ,
\end{eqnarray}

\noindent where the relation between the adiabatic phases
$\gamma_j (t)$ and $\tilde{\gamma}_j (t)$ is

\begin{eqnarray} \label{9}
\tilde{\gamma}_j (t) = \gamma_j (t) - \alpha_j (t) + \alpha_j (0),
     \hspace{1cm} j= 1, 2, \cdots, M.
\end{eqnarray}

The density of probability written in the new basis is,

\begin{eqnarray}  \label{10}
|\psi (\vec{x}, t)|^2 = &=& \sum_{j=1}^{M}  |\tilde{a}_j|^2 |\Phi_j (\vec{x}; t)|^2
        \nonumber  \\
%
%
&+&  \sum_{{j, l=1}  \atop {j \not= l}}^{M} \tilde{a}_j \, \tilde{a}_l^{*} 
  e^{ i [\tilde{\gamma}_j (t) - \tilde{\gamma}_l (t)]} \;\; 
    e^{- \frac{it}{\hbar} \left[\langle E_j(t) \rangle -  \langle E_l(t) \rangle\right]} \;\;
       \Phi_j (\vec{x};t) \Phi_l^{*} (\vec{x};t) .
\end{eqnarray}

From the eqs. (\ref{6}), (\ref{7b}) and (\ref{9}), we obtain

\begin{eqnarray}  \label{11}
\tilde{a}_j \, \tilde{a}_l^{*} 
  e^{ i [\tilde{\gamma}_j (t) - \tilde{\gamma}_l (t)]} \;\; 
       \Phi_j (\vec{x};t) \Phi_l^{*} (\vec{x};t)  =
a_j \, a_l^{*}  e^{ i [\gamma_j (t) - \gamma_l (t)]} \;\; 
       \varphi_j (\vec{x};t) \varphi_l^{*} (\vec{x};t), 
\end{eqnarray}

\noindent $l, j = 1, 2, \cdots, M$. In eq.(\ref{11}) we
include the terms $l=j$. Result (\ref{11}) tells us that 
each term in the two sums on the r.h.s. of eq.(\ref{5}) is
independent of the basis of the instantaneous eigenvectors
of energy that we use to do the calculation. 
We point out that to obtain the corrected
result for $|\psi (\vec{x}, t)|^2$ we have to take 
into account the difference of the adiabatic phases
(\ref{3a}), that are non-physical.

\vspace{0.5cm}

To generalize the conclusions derived from result (\ref{11})
we consider {\bf O} to be an hermitian operator associated 
to a physical quantity. The time-evolution of the average
of this operator in the initial state (\ref{2}) is

\begin{eqnarray}  \label{12}
  o (t) &=&  \langle \psi (t)| {\bf O} | \psi(t)\rangle
               \nonumber \\
%
%
&=& \sum_{j=1}^{M} |a_j|^2 \; \langle \varphi_j; t| {\bf O}| \varphi_j; t\rangle
+ \sum_{{j, l=1}  \atop {j \not= l}}^{M} a_j \, a_l^{*} 
  e^{ i [\gamma_j (t) - \gamma_l (t)]} \;\; 
    e^{- \frac{it}{\hbar} \left[\langle E_j(t) \rangle -  \langle E_l(t) \rangle\right]} \;\;
       \langle \varphi_j; t| {\bf O}| \varphi_l; t\rangle .   \nonumber\\
%
%
&&
\end{eqnarray}

Following the same steps as we did to proof that the 
terms that contribute to the density probability
is basis-independent, we show that the same is true 
for each term in the two sums on the r.h.s. of
eq.(\ref{12}).

Therefore  if we use a basis of instantaneous 
eigenstates of energy, that is not of parallel 
transported states, to describe the adiabatic evolution 
of a vector state that initially is in a superposition of 
eigenstates of energy at $t=0$, the non-physical 
adiabatic  phases  (\ref{3a}) have to be taken into 
account to give the correct interference
terms when we calculate physical quantities.

\vspace{0.5cm}

To exemplify the importance to take into account 
the adiabatic phases (\ref{3b}) to obtain 
the correct result in physical quantities, we 
consider the soluble model of the spin-$1/2$ 
in the presence of an external classical 
magnetic field. This field precesses around a 
$z$-direction with constant angular frequency 
$\omega_0$. This model was discussed by Berry in  
Ref. \cite{berry} and by Garc\'ia de Polavieja and Sj\"oqvist
in Ref.\cite{gonzalo}. Being a soluble model
we can verify the result obtained in the adiabatic 
regime by applying the adiabatic approximation directly 
in the exact result\cite{ajp2000}.

The hamiltonian of a spin-$1/2$ in the presence of an 
external classical magnetic field $\vec{B} (t)$ is\cite{ajp2000}

\begin{subequations}

\begin{eqnarray}   \label{13}
{\bf H} (t) = \frac{\mu \hbar}{2} \vec{B} (t)\cdot \vec{\sigma},
\end{eqnarray}

\noindent where 

\begin{eqnarray}  \label{14}
\vec{B} (t) = ( B \sin(\theta) \cos(\omega_0 t), B\sin(\theta) \sin(\omega_0 t), B\cos(\theta)), 
\end{eqnarray}

\noindent  with $B\equiv |\vec{B}|$ and $\theta$ is the 
angle between the  external magnetic field and  the 
$z$-direction. The $\sigma_i$, $i \in \{x, y, z\}$ 
are the Pauli matrices, $\mu= g \mu_B$, where 
$\mu_B$ is the Bohr magneton and $g$ is the 
Land\'e's factor.

\end{subequations}

In Ref.\cite{ajp2000} we obtain the two eigenvectors
of hamiltonian (\ref{13}) and their respective eigenvalues,

\begin{subequations}

\begin{eqnarray} \label{15}
|\phi_1; t\rangle = -\sin(\frac{\theta}{2}) |\uparrow\rangle 
+ \cos(\frac{\theta}{2}) e^{i \omega_0 t} |\downarrow\rangle
   \hspace{0.5cm} &\Rightarrow&  \hspace{0.5cm} 
     E_1 = -\frac{\mu \hbar B}{2} ,
                  \label{15a} \\
\nonumber \\
%
%
|\phi_2; t\rangle = \cos(\frac{\theta}{2}) |\uparrow\rangle 
+ \sin(\frac{\theta}{2}) e^{i \omega_0 t} |\downarrow\rangle
\hspace{0.5cm} &\Rightarrow& \hspace{0.5cm} E_2 = \frac{\mu \hbar B}{2} .
                  \label{15b} 
\end{eqnarray}

\end{subequations}

\noindent We denote the eigenvector of $\sigma_z$ with 
eigenvalue +1 (-1) to be
$|\uparrow\rangle$ ($|\downarrow\rangle$).

We choose the initial vector state of the spin-$1/2$ 
system to be,

\begin{eqnarray} \label{16}
|\psi (0)\rangle = a_1 | \phi_1; 0\rangle + a_2 |\phi_2; 0\rangle
\end{eqnarray}

\noindent and $|a_1|^2 +  |a_2|^2 =1$. For simplicity we 
take $a_1$ and $a_2 \in \mathbb{R}$.

From eq.(\ref{3a}), the adiabatic evolution of the previous
initial state is

\begin{eqnarray} \label{17}
|\psi (t) \rangle = e^{i\gamma_1 (t)} \, e^{-\frac{it E_1}{\hbar}}
\;\; \left[ a_1 |\phi_1; t\rangle + a_2 \; e^{i \alpha [\gamma_2 (t) - \gamma_1 (t)]}
   \; e^{-\frac{it}{\hbar} [E_2 - E_1]} \;\; |\phi_2; t\rangle \right].
\end{eqnarray}

The previous equation is similar to eq.(46) of Ref.\cite{gonzalo}.

From a direct calculation of phase (\ref{3b}), we obtain: 
$\gamma_1 (t) = - \frac{(1+ \cos(\theta)) \omega_0 t}{2}$  and
$\gamma_2 (t) = - \frac{(1- \cos(\theta)) \omega_0 t}{2}$.

In eq.(\ref{17}) we include the tracer $\alpha$ to verify 
if the difference of adiabatic phases contribute
to physical quantities. At the end of the calculation
we take $\alpha = 1$.

The expectation value of the operator ${\bf s}_z$ in the
state $|\psi (t)\rangle$ is

\begin{eqnarray}  \label{18}
\langle \psi (t)| {\bf s}_z| \psi (t)\rangle = \frac{\hbar}{2} \cos (\theta) [a_2^2 - a_1^2]
- 4 a_1 a_2 \hbar \sin (\theta)  \cos[(\mu B - \alpha \omega_0 \cos(\theta)) t].
\end{eqnarray}

From what we discussed in the first part of this letter,
result (\ref{18}) is physical. We verify that 
the adiabatic phases (\ref{3b}) contribute to the second 
term on the r.h.s. of expression (\ref{18}) with
a phase that increases in time up to reach the difference 
of Berry's phases when the $t= \frac{2\pi}{\omega_0}$.

In Ref. \cite{ajp2000} we have the exact dynamics of the initial vector
(\ref{16}). Using the exact time dependence of the $|\psi (t)\rangle$ we 
calculate the expectation value of the operator ${\bf s}_z$ and
implement in it the adiabatic approximation. This approximated
result coincides with expression (\ref{18}) with $\alpha =1$.

\vspace{0.5cm}

In conclusion, we show that although the adiabatic
phase (\ref{3b}) is non-physical, the phase differences
do contribute to physical quantities during 
the adiabatic evolution of  a noncyclic quantum system 
if the vector state is  a superposition of instantaneous 
eigenstates of energy. 

The result  of each interference term on the 
r.h.s. of eq.(\ref{12}) is independent of a particular
choice of  basis of the  instantaneous energy eigenstates.
In order to verify the consequences of an adiabatic
variation of the hamiltonian on the motion of the 
quantum system driven by it, we do not need to
define a noncyclic geometric phase, as it has been done 
in the literature.

In order to show the importance of the contribution 
of this  phase difference of adiabatic nature,
we calculate the expectation value of the operator
${\bf s}_z$ of a spin-$1/2$ model coupled
to an external magnetic field that precesses around 
a fixed  direction. We verify that the adiabatic
approximation of
$\langle \psi (t)| {\bf s}_z| \psi(t)\rangle$,
derived from its exact expression, only coincides 
with the calculation of the adiabatic evolution
of this operator if the difference of the adiabatic
phases (\ref{3b}) is included in the dynamics of the 
instantaneous eigenstates (\ref{15a}) and (\ref{15b}).

\vspace{1cm}

M.T. Thomaz  (Fellowship CNPq, Brazil, Proc.No.: 30.0549/83-FA) 
thanks CNPq  for partial financial support and to the Univ. Estadual 
do Mato Grosso do Sul, campus of Dourados, where part of this
work was realized.


\end{document}